\documentclass[10pt]{revtex4-2}
\usepackage[T1]{fontenc}
\usepackage[utf8]{inputenc}
\usepackage{fancybox}
\usepackage{calc}
\usepackage{amsmath}
\usepackage{amssymb}
\usepackage{stmaryrd}
\PassOptionsToPackage{version=3}{mhchem}
\usepackage{mhchem}

\makeatletter
\usepackage{amsfonts}
\usepackage[export]{adjustbox}
\graphicspath{ {./images/} }

\makeatother

\begin{document} 
\title{Vanishing of Conserved Charges in Cotton Gravity}
\author{Emel Altas}
\email{emelaltas@kmu.edu.tr}

\affiliation{Department of Physics,~\\
 Karamanoglu Mehmetbey University, 70100, Karaman, Turkey}
\author{Bayram Tekin\thanks{Corresponding author} }
\email{btekin@metu.edu.tr}

\affiliation{Department of Physics,~\\
 Middle East Technical University, 06800, Ankara, Turkey}
\date{\today}
\begin{abstract}
\noindent Cotton gravity was recently introduced as a higher derivative extension of General Relativity. The field equations of the theory involve the rank-3 Cotton tensor. Here we show that all solutions of the theory, including the black holes, have vanishing conserved charges, {\it i.e.} mass and angular momentum. This result implies that either the theory is unphysical since all the black holes carry the charges of the vacuum and can be created at no energy cost, or the theory has confinement of mass/energy and all other conserved quantities.
\end{abstract}
\maketitle

\section{Introduction}
Modifications of Einstein's gravity are motivated for well-known reasons such as doing away with the need for dark energy and dark matter, or short distance problems such as the singularities in cosmology and the black hole physics \cite{metric}. Due to similar motivations, 
in \cite{Harada} the rank-3 Cotton tensor, which involves the third derivatives of the metric tensor, was used to define the following modified gravity theory 
\begin{equation}
C^{\nu\rho\sigma}=8\pi GT^{\nu\rho\sigma},\label{denklem1}
\end{equation}
where the $n-$dimensional Cotton tensor is defined as 
\begin{equation}
C_{\nu\rho\sigma}:=\nabla_{\rho}R_{\nu\sigma}-\nabla_{\sigma}R_{\nu\rho}-\frac{1}{2(n-1)}\left(g_{\nu\sigma}\nabla_{\rho}R-g_{\nu\rho}\nabla_{\sigma}R\right).\label{denklem2}
\end{equation}
Here $R_{\mu \nu}$ is the Ricci tensor and $R$ is the scalar curvature. The right-hand side of (\ref{denklem1}), that is the matter content, involves the following tensor
\begin{equation}
T^{\nu\rho\sigma}:=\nabla^{\rho}T^{\nu\sigma}-\nabla^{\sigma}T^{\nu\rho}-\frac{1}{n-1}\left(g^{\nu\sigma}\nabla^{\rho}T-g^{\nu\rho}\nabla^{\sigma}T\right),
\end{equation}
where $T^{\mu \nu}$ is the covariantly conserved energy-momentum tensor and $T$ is its trace. In $n \ge 4$, dimensions one can prove that the Weyl tensor 
 \begin{eqnarray}
      W_{\mu \nu \rho \sigma} := R_{\mu\nu\rho \sigma}+{\frac {1}{n-2}}\left(R_{\mu \sigma}g_{\nu \rho }-R_{\mu \rho }g_{\nu \sigma}+R_{\nu \rho }g_{\mu \sigma}-R_{\nu \sigma}g_{\mu \rho }\right)
      +{\frac {1}{(n-1)(n-2)}}R\left(g_{\mu \rho }g_{\nu \sigma}-g_{\mu \sigma}g_{\nu \rho }\right), \label{denklem3}
 \end{eqnarray}
is a {\it potential } for the Cotton tensor, namely one has the identity for all smooth metrics:
\begin{equation}
\frac{n-3}{n-2} C_{\nu\rho\sigma} = \nabla_\mu W^\mu\,_{\nu \rho \sigma}, \hskip 1 cm n \ge 4. \label{denklem4}
\end{equation}
So clearly, from this expression, one can conclude that conformally flat metrics are Cotton-flat and constitute a set of vacuum solutions of the theory. However, some Cotton-flat metrics are not Weyl-flat.
 Interestingly, in 4 dimensions, there is a rank-3 tensor, the so-called Lanczos potential for the Weyl tensor, and in $n>4$ dimensions, there is a rank-5 potential for the Weyl tensor \cite{Edgar:2004iq}. These potentials can be used to explore the solution space of the theory. For example, using (\ref{denklem4}), one can recast the Cotton gravity equations as follows: One has 
\begin{equation}
T^{\nu\rho\sigma} = \nabla_\mu \Theta^{\mu \nu \rho \sigma},
\end{equation}
where the potential for the rank-3 energy momentum tensor is a rank-4 tensor:
\begin{equation}
\Theta^{\mu \nu \rho \sigma}:= g^{ \mu \rho}T^{\nu\sigma}-g^{\mu \sigma}T^{\nu\rho}-\frac{1}{n-1}\left(g^{\nu\sigma}g^{ \mu \rho}-g^{\nu\rho}g^{ \mu \sigma}\right) T.
\end{equation}
Therefore, the field equations of Cotton gravity read as
\begin{equation}
\nabla_\mu \Bigg( \frac{n-2}{n-3} W^{\mu \nu \rho \sigma}-8\pi G \Theta^{\mu \nu \rho \sigma} \Bigg)=0.
\end{equation}

Cotton gravity received some attention in the literature about its reformulation, validity, exact solutions, etc. \cite{Bargueno,Harada2,Mantica,Gog,Mantica2,Clement,Sussman,Junior,Gurses}. Here we will construct the conserved gravitational charges of the theory based on Killing symmetries following the methods introduced in \cite{AD,DT1,DT2}, and summarized in \cite{Adami}. Our result is disparaging for this alternative gravity theory: The conserved mass/energy corresponding to the (asymptotically) timelike Killing vector vanishes identically, and the angular momentum corresponding to the rotational Killing vector also vanishes, as long as there is no matter field at spatial infinity. Creating a black hole does not cost energy and angular momentum in the theory, as the black hole has zero energy and angular momentum. This signals an unavoidable instability as black holes will pop up from the vacuum to increase the total entropy of the universe without costing any energy or angular momentum. Another interpretation would be that the theory is strongly interacting and all mass and angular momentum are confined in a region which was first advocated in \cite{BS} for the Weyl-square theory in four dimensions.

\section{Conserved Charges of Cotton gravity}

\subsection{Constructing the conserved current}

To construct the conserved charges of Cotton gravity, let us recall some of the properties of the Cotton tensor \cite{Garcia}.
\begin{enumerate}
\item Cotton tensor is divergence-free with respect to its first index 
\begin{equation}
\nabla_{\mu}C^{\mu\sigma\rho}=0,
\end{equation}
\item it is antisymmetric in its last two indices
\begin{equation}
C^{\mu\sigma\rho}=-C^{\mu\rho\sigma},
\end{equation}
\item it satisfies an algebraic Bianchi-type identity 
\begin{equation}
C^{\mu\rho\sigma}+C^{\rho\sigma\mu}+C^{\sigma\mu\rho}=0.
\end{equation}
\end{enumerate}
Taking the divergence of the last equation, one arrives at
\begin{equation}
\nabla_{\mu}C^{\mu\rho\sigma}+\nabla_{\mu}C^{\rho\sigma\mu}+\nabla_{\mu}C^{\sigma\mu\rho}=0,
\end{equation}
which reduces to
\begin{equation}
\nabla_{\mu}C^{\rho\sigma\mu}=\nabla_{\mu}C^{\sigma\rho\mu}.
\end{equation}
So, in this form, when one takes the covariant divergence with respect to the third index, one obtains a symmetric rank-2 tensor in the first two indices.
 We will now employ this fact to construct a conserved current for a given Killing vector. Let $\xi^\mu$ be a Killing vector, then it satisfies
\begin{equation}
\nabla_{\mu}\nabla_{\rho}\xi_{\sigma}=R^{\lambda}{}_{\mu\rho\sigma}\xi_{\lambda}.
\end{equation}
Using the above identities on the Cotton tensor and the Killing vector, one arrives at the following covariantly conserved current
\begin{equation}
J^{\mu}=C^{\mu\rho\sigma}\nabla_{\rho}\xi_{\sigma}-2\xi_{\sigma}\nabla_{\rho}C^{\sigma\mu\rho}. \label{current1}
\end{equation}
Namely one has $\nabla_\mu J^\mu =0$. The second step in the conserved charge construction is to show that the current $J^\mu$ can be written as the divergence of an antisymmetric rank-2 tensor. This step involves a slightly longer calculation which we give in the appendix and quote the result here. One finds that the full current can be written as 
\begin{equation}
J^{\mu}=\nabla_{\sigma}\left(2\xi_{\lambda}C^{\lambda\sigma\mu}\right). \label{full0}
\end{equation}
Note that the factor 2 is not important, but we keep it to retain consistency with the other equations. Using the field equations (\ref{denklem1}), we then have 
\begin{equation}
  J^\sigma:= \nabla_\rho \left (2\xi_\nu C^{\nu\rho\sigma} \right)=16\pi G  \nabla_\rho \left (\xi_\nu T^{\nu\rho\sigma} \right).
\end{equation}
Already at this stage, an astute reader will recognize that the conserved current structure of this theory signals trouble in the sense that the current is obtained from the divergence of the Cotton tensor, i.e. the tensor used in defining the field equations. So in a vacuum, far outside the sources, the current is identically zero and this will yield zero conserved charges for all solutions. Recall that in Einstein's theory the situation is quite different: one has $J^\mu = \xi_\nu G^{\nu \mu} = \nabla_\nu {\mathcal{F}^{\nu \mu}(\xi,h)}$ where $\mathcal{F}^{\nu \mu}$ is antisymmetric and given explicitly in terms of the deviation of the metric ($h$) from the vacuum and the Killing vector. See equations (28) and (29) in \cite{Adami}.

Let us explicitly show what is stated in the previous paragraph. There are two minor issues we still have to fix: 
Firstly, as $J^{\mu}$ is {\it covariantly} conserved, we can build a {\it partially} conserved current density 
$\mathcal{J}^\mu:= \sqrt{-g}J^{\mu}$,  such that $\partial_\mu \mathcal{J}^\mu=0$  via $\sqrt{-g} \nabla_\mu J^\mu = \partial_{\mu} \left( \sqrt{-g} J^\mu \right)=0$. Secondly, we must define the conserved current and charges of the vacuum to be zero. The vacuum must at least satisfy the matter-free equation, i.e. $\bar C^{\nu\rho\sigma}=0$. Of course, there are many other solutions to this equation, such as black holes, but at this stage let $\bar {g}_{\mu \nu}$ denote the vacuum with at least one Killing vector. Then, by construction 
$\mathcal{\bar J}^\mu=0$ and linearization around the vacuum yields
\begin{equation}
\mathcal{J}^\mu= \sqrt{-\bar g}\bar\nabla_{\sigma}\left(2\bar \xi_{\lambda}C_{(L)}^{\lambda\sigma\mu}\right). \label{lin1}
\end{equation}
Following the steps given in Section III of \cite{Tekin} verbatim, one can build the conserved charges for each Killing vector field.
 To do so, let $\bar {g}_{\mu \nu}$ be the background metric on the manifold that is considered to be the vacuum with zero conserved charges as noted above. We assume that there is at least one Killing vector $\bar{\xi}^\mu$ that the vacuum possesses, and let $\bar{\mathcal{M}}$ denote this geometry and $h_{\mu \nu}$ denote a symmetric field living on it in such a way that it decays sufficiently fast to make the relevant integrals finite. Then integrating $\partial_\mu \mathcal{J}^\mu =0$ over the manifold, and using Stokes' theorem, one has  
\begin{equation}
0= \int_{\bar{\mathcal{M}}} d^n x \,\partial_\mu \mathcal{J}^\mu = \int_{\partial {\bar{\mathcal{M}} }} d^{n-1}y\, {\hat n}_\mu \mathcal{J}^\mu, 
\label{equal-zero}
\end{equation}
Here $\partial {\bar{\mathcal{M}} }$ is the total non-null boundary of the spacetime, and ${\hat n}_\mu $ is a unit normal time-like co-vector to the boundary as depicted in Figure 1. The side parts of this ``cylindrical`` spacetime constitute a time-like boundary at spatial infinity where we assume there is no current.
\begin{figure}
\begin{centering}
\includegraphics[scale=0.5]{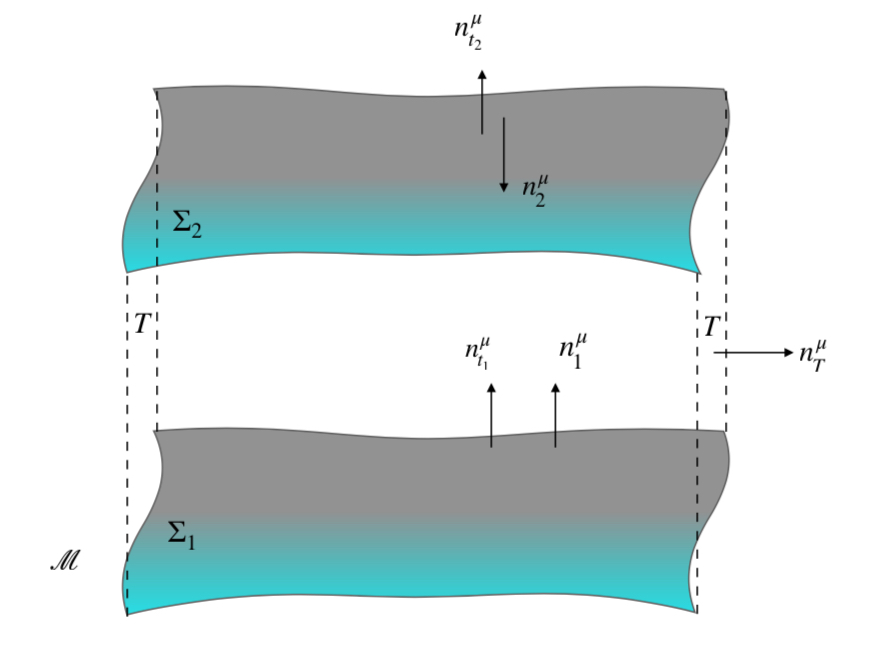} 
\par\end{centering}
\caption{Here $\Sigma_{1}$ and $\Sigma_{2}$ and $T$ are co-dimension-one hypersurfaces that constitute the boundary of spacetime. $\Sigma_{1}$ and $\Sigma_{2}$  are spacelike with time-like normal vectors. The timelike surface $T$ is at spatial infinity with no flux. Note that we took the figure from \cite{Tekin}. }
\end{figure}
To get a non-zero result, we define the
integration only over a spatial hypersurface $\bar{\Sigma}_1$ which is not equal to the total boundary $\partial {\bar{\mathcal{M}} }$ as
\begin{equation}
Q[\bar \xi, \bar{\Sigma}_1 ]:= \int_{\bar{\Sigma}_1 } d^{n-1}y\, {\hat n}_\mu \mathcal{J}^\mu =\int_{\bar{\Sigma}_1 } d^{n-1}y\
\sqrt{\bar{\gamma}}\, {\hat n}_\mu \bar\nabla_{\sigma}\left(2\bar \xi_{\lambda}C_{(L)}^{\lambda\sigma\mu}\right),
\label{charge1}
\end{equation}
where $\gamma_{\mu \nu}$ is the pullback metric on $\bar{\Sigma}_1$ and in the second equality we used  (\ref{lin1}). As is clear from Figure 1, $\partial {\bar{\mathcal{M}} } =  \bar{\Sigma}_1 \cup \bar{\Sigma}_2\cup \bar T $,  then the statement (\ref{equal-zero}) can be restated as a charge-conservation statement as
\begin{equation}
Q[\bar \xi, \bar{\Sigma}_1 ] =Q[\bar \xi, \bar{\Sigma}_2 ].
\end{equation} 
which means we can use any $\bar \Sigma_t$ at time $t$, hence the name conserved charge. In (\ref{charge1}), we can apply Stokes' theorem once again to get 

\begin{equation}
Q[\bar \xi] = 2 \int_{\partial \bar{\Sigma}} d^{n-2}z\
\sqrt{\bar{q}}\, {\hat n}_\mu \hat \sigma_\nu\bar \xi_{\lambda}C_{(L)}^{\lambda\nu\mu},
\label{charge2}
\end{equation}
where $\bar q_{\mu \nu}$ is the induced metric on $\partial \bar{\Sigma}$; and  $\hat \sigma_\nu$ is the outward unit normal co-vector on it.   Note that since $\hat n_\mu$ is of the form $(1,0,0,0..)$,  $\nabla_\nu \hat n_\mu$ is symmetric and its contraction with the antisymmetric tensor
$(\bar \xi_{\lambda}C_{(L)}^{\lambda\nu\mu})$ is zero. We can rewrite the final result as 

 \begin{equation}
Q[\bar \xi] = 2\int_{\partial \bar{\Sigma}} d^{n-2}z\
\sqrt{\bar{q}}\, \bar{\epsilon}_{\mu \nu} \bar \xi_{\lambda}C_{(L)}^{\lambda\nu\mu},  \hskip 1 cm \bar{\epsilon}_{\mu \nu} :=\frac{1}{2}\left (\hat n_\mu \hat \sigma_\nu -{\hat n}_\nu \hat \sigma_\mu \right ),
\label{charge3}
\end{equation}
where the Linearized Cotton tensor is
\begin{align}
\left(C^{\nu\rho\sigma}\right)^{(L)} & =\bar{\nabla}^{\rho}\left(R^{\nu\sigma}\right)^{(L)}-\bar{\nabla}^{\sigma}\left(R^{\nu\rho}\right)^{(L)}-\frac{1}{2(n-1)}\left(\bar{g}^{\nu\sigma}\bar{\nabla}^{\rho}(R)^{(L)}-\bar{g}^{\nu\rho}\bar{\nabla}^{\sigma}(R)^{(L)}\right).
\end{align}
Assuming that there is no matter at $\partial \bar{\Sigma}$, we have $C^{\mu \nu \sigma}=0$ there, and hence its linearization is also zero. Therefore, for all solutions $Q[\bar \xi] =0$, and in particular for the time-like  Killing vector $\bar \xi = \frac{\partial}{\partial t}$, the corresponding conserved charge mass/energy vanishes. 

\section{CONCLUSIONS}

We have shown that Cotton gravity theory has identically zero conserved charges for all of its solutions as long as there is no matter at the spatial boundary of spacetime. This shows that the theory is unphysical in the sense that a vacuum without a black hole is degenerate with a space-time with many black holes. Hence, the theory does not have a viable ground state. For example, the parameter $m$ in the spherically symmetric solution given in \cite{Harada} cannot be interpreted as the conserved mass. Let us also note that there could be another, albeit unlikely, explanation for this vanishing conserved-charge result: In pure Weyl square theory in the metric formulation, a zero energy theorem for all solutions was interpreted as the confinement of energy in \cite{BS}. This was later reanalyzed and disputed within the Killing charge construction in \cite{DT2} where a section was devoted to possible zero-energy models. Weyl gravity and Cotton gravity are related in the sense that the former involves the usual metric variation of the Weyl square action, while the latter involves a variation with respect to the Christoffel connection \cite{Harada}.

\begin{acknowledgments}
We dedicate this work to the memory of Prof. Dr. I. Ferit Oktem (1928-2024), one of the first researchers to write papers on relativity from Turkey. The authors thank Metin Gurses for useful discussions.
E. Altas is supported by TUBITAK Grant No. 123F353.
\end{acknowledgments}
\section{Appendix: Conserved current as a divergence of an antisymmetric tensor}

Let us start with the current given in equation (\ref{current1}).
 We want to express the current as a divergence, such that
\begin{equation}
J^{\mu}=\nabla_{\sigma}X^{\sigma\mu},
\end{equation}
where $X^{\sigma\mu}$ is expected to be antisymmetric. Since the Cotton tensor is antisymmetric in its last two indices, the last term in equation (\ref{current1}) is already in the desired form. Moreover,
for the second term on the right hand side of equation (\ref{current1}),
we can use the identity
\begin{equation}
C^{\sigma\mu\lambda}+C^{\mu\lambda\sigma}+C^{\lambda\sigma\mu}=0,
\end{equation}
and so
\begin{equation}
C^{\sigma\mu\lambda}=-C^{\mu\lambda\sigma}-C^{\lambda\sigma\mu}.
\end{equation}
Then $J^{\mu}$ becomes
\begin{equation}
J^{\mu}=C^{\mu\sigma\lambda}\nabla_{\sigma}\xi_{\lambda}-2\nabla_{\sigma}\Bigl(\xi_{\lambda}\left(-C^{\mu\lambda\sigma}-C^{\lambda\sigma\mu}+C^{\lambda\mu\sigma}\right)\Bigr).
\end{equation}
Using $\quad C^{\lambda\sigma\mu}=-C^{\lambda\mu\sigma}$, we obtain
\begin{equation}
J^{\mu}=C^{\mu\sigma\lambda}\nabla_{\sigma}\xi_{\lambda}-2\nabla_{\sigma}\Bigl(\xi_{\lambda}\left(-C^{\mu\lambda\sigma}+C^{\lambda\mu\sigma}+C^{\lambda\mu\sigma}\right)\Bigr),
\end{equation}
and
\begin{equation}
\begin{aligned}J^{\mu}=C^{\mu\sigma\lambda}\nabla_{\sigma}\xi_{\lambda}-2\nabla_{\sigma}\Bigl(\xi_{\lambda}\left(-C^{\mu\lambda\sigma}+2C^{\lambda\mu\sigma}\right)\Bigr).\end{aligned}
\end{equation}
We re-express the conserved current as
\begin{equation}
J^{\mu}=C^{\mu\sigma\lambda}\nabla_{\sigma}\xi_{\lambda}+2\nabla_{\sigma}\left(\xi_{\lambda}C^{\mu\lambda\sigma}\right)-4\nabla_{\sigma}\left(\xi_{\lambda}C^{\lambda\mu\sigma}\right).
\end{equation}
Let us focus on the first two terms on the right hand side: for simplicity
introduce a new tensor field
\[
X^{\mu}:=C^{\mu\sigma\lambda}\nabla_{\sigma}\xi_{\lambda}+2\nabla_{\sigma}\left(\xi_{\lambda}C^{\mu\lambda\sigma}\right).
\]
And so we write
\begin{equation}
J^{\mu}=X^{\mu}-4\nabla_{\sigma}\left(\xi_{\lambda}C^{\lambda\mu\sigma}\right).\label{eq:cc}
\end{equation}
More explicitly it becomes
\begin{equation}
X^{\mu}=C^{\mu\sigma\lambda}\nabla_{\sigma}\xi_{\lambda}+2\nabla_{\sigma}\xi_{\lambda}C^{\mu\lambda\sigma}+2\xi_{\lambda}\nabla_{\sigma}C^{\mu\lambda\sigma},
\end{equation}
and
\begin{equation}
X^{\mu}=\nabla_{\sigma}\xi_{\lambda}\left(C^{\mu\sigma\lambda}+2C^{\mu\lambda\sigma}\right)+2\xi_{\lambda}\nabla_{\sigma}C^{\mu\lambda\sigma},
\end{equation}
which yields the followings
\begin{align}
X^{\mu} & =\nabla_{\sigma}\xi_{\lambda}\left(C^{\mu\sigma\lambda}-2C^{\mu\sigma\lambda}\right)+2\xi_{\lambda}\nabla_{\sigma}C^{\mu\lambda\sigma}\nonumber \\
 & =-\left(\nabla_{\sigma}\xi_{\lambda}\right)C^{\mu\sigma\lambda}+2\xi_{\lambda}\nabla_{\sigma}C^{\mu\lambda\sigma}.
\end{align}
Use the identity $C^{\mu\lambda\sigma}+C^{\lambda\sigma\mu}+C^{\sigma\mu\lambda}=0$,
to arrive at
\begin{equation}
X^{\mu}=-\left(\nabla_{\sigma}\xi_{\lambda}\right)C^{\mu\sigma\lambda}-2\xi_{\lambda}\nabla_{\sigma}\left(C^{\lambda\sigma\mu}+C^{\sigma\mu\lambda}\right)=0,
\end{equation}
where $\nabla_{\sigma}C^{\sigma\mu\lambda}=0$, and so one has
\begin{equation}
X^{\mu}=-\left(\nabla_{\sigma}\xi_{\lambda}\right)C^{\mu\sigma\lambda}-2\xi_{\lambda}\nabla_{\sigma}C^{\lambda\sigma\mu}.
\end{equation}
Recall that $C^{\lambda\sigma\mu}$ has the desired antisymmetry.
Therefore, we can express this piece as a divergence. We have
\begin{equation}
\begin{aligned}X^{\mu} & =-\left(\nabla_{\sigma}\xi_{\lambda}\right)C^{\mu\sigma\lambda}-2\xi_{\lambda}\nabla_{\sigma}C^{\lambda\sigma\mu}\\
 & =-\left(\nabla_{\sigma}\xi_{\lambda}\right)C^{\mu\sigma\lambda}-2\nabla_{\sigma}\left(\xi_{\lambda}C^{\lambda\sigma\mu}\right)+2\left(\nabla_{\sigma}\xi_{\lambda}\right)C^{\lambda\sigma\mu},
\end{aligned}
\end{equation}
which yields
\begin{equation}
X^{\mu}=\nabla_{\sigma}\xi_{\lambda}\left(-C^{\mu\sigma\lambda}+2C^{\lambda\sigma\mu}\right)-2\nabla_{\sigma}\left(\xi_{\lambda}C^{\lambda\sigma\mu}\right)
\end{equation}
or 
\begin{equation}
X^{\mu}=\nabla_{\sigma}\xi_{\lambda}\left(C^{\sigma\lambda\mu_{+}}C^{\lambda\mu\sigma}+2C^{\lambda\sigma\mu}\right)-2\nabla_{\sigma}\left(\xi_{\lambda}C^{\lambda\sigma\mu}\right).
\end{equation}
Clearly one obtains
\begin{equation}
X^{\mu}=\nabla_{\sigma}\xi_{\lambda}\left(C^{\sigma\lambda\mu}-C^{\lambda\mu\sigma}\right)-2\nabla_{\sigma}\left(\xi_{\lambda}C^{\lambda\sigma\mu}\right),
\end{equation}
and more explicitly
\begin{equation}
X^{\mu}=\left(\nabla_{\sigma}\xi_{\lambda}\right)C^{\sigma\lambda\mu}-\left(\nabla_{\sigma}\xi{}_{\lambda}\right)C^{\lambda\mu\sigma}-2\nabla_{\sigma}\left(\xi_{\lambda}C^{\lambda\sigma\mu}\right).
\end{equation}
Since the Cotton tensor is divergence free with respect to its first
index, the first term on the right hand side can be written as a total
derivative:
\begin{equation}
\left(\nabla_{\sigma}\xi_{\lambda}\right)C^{\sigma\lambda\mu}=\nabla_{\sigma}\left(\xi_{\lambda}C^{\sigma\lambda\mu}\right),
\end{equation}
and for the second term we should use the Killing equation, $\nabla_{\sigma}\xi{}_{\lambda}=-\nabla_{\lambda}\xi_{\sigma}$,
to obtain
\begin{equation}
\left(\nabla_{\sigma}\xi_{\lambda}\right)C^{\lambda\mu\sigma}=-\left(\nabla_{\lambda}\xi_{\sigma}\right)C^{\lambda\mu\sigma}=-\nabla_{\lambda}\left(\xi_{\sigma}C^{\lambda\mu\sigma}\right).
\end{equation}
Finally, one has
\begin{equation}
X^{\mu}=\nabla_{\sigma}\left(\xi_{\lambda}C^{\sigma\lambda\mu}\right)+\nabla_{\lambda}\left(\xi_{\sigma}C^{\lambda\mu\sigma}\right)-2\nabla_{\sigma}\left(\xi_{\lambda}C^{\lambda\sigma\mu}\right).
\end{equation}
Renaming the indices, we obtain
\begin{equation}
X^{\mu}=\nabla_{\sigma}\Bigl(\xi_{\lambda}\left(C^{\sigma\lambda\mu}+C^{\sigma\mu\lambda}-2C^{\lambda\sigma\mu}\right)\Bigr),
\end{equation}
where the first two terms cancel each other. We then end up with
\begin{equation}
X^{\mu}=-2\nabla_{\sigma}\left(\xi_{\lambda}C^{\lambda\sigma\mu}\right).
\end{equation}
Inserting our result in (\ref{eq:cc}), we arrive at

\begin{align}
J^{\mu} & =-2\nabla_{\sigma}\left(\xi_{\lambda}C^{\lambda\sigma\mu}\right)-4\nabla_{\sigma}\left(\xi_{\lambda}C^{\lambda\mu\sigma}\right)\nonumber \\
 & =\nabla_{\sigma}\left(\xi_{\lambda}\left(-2C^{\lambda\sigma\mu}-4C^{\lambda\mu\sigma}\right)\right)\\
 & =\nabla_{\sigma}\left(\xi_{\lambda}\left(-2C^{\lambda\sigma\mu}+4C^{\lambda\sigma\mu}\right)\right),\nonumber 
\end{align}
and so
\begin{equation}
J^{\mu}=\nabla_{\sigma}\left(2\xi_{\lambda}C^{\lambda\sigma\mu}\right),
\end{equation}
which is the result we were supposed to find.

\end{document}